%% file: paper.tex
\documentclass[runningheads,a4paper]{llncs}

\usepackage[latin1]{inputenc}
\usepackage{amsmath}
\usepackage{listings}
\usepackage{marvosym}
\usepackage{color}
\usepackage{graphicx}
\usepackage{subfigure}
\usepackage{caption}
\usepackage{listings}

\usepackage{pgf}
\usepackage{tikz}
\usetikzlibrary{arrows,automata,positioning}
\usepackage{epstopdf}
\newtheorem{mydef}{Definition}
\newtheorem{mylem}{Lemma}

\newcommand{\comment}[1]{}

\usepackage{listings}
\usepackage{amssymb}
\usepackage{color}
\usepackage{graphicx}

\usepackage{url}

\begin{document}

\mainmatter  % start of an individual contribution

% first the title is needed
\title{Model Checking C Programs with Loops via \textit{k}-Induction and Invariants}

% a short form should be given in case it is too long for the running head
% \titlerunning{Lecture Notes in Computer Science: Authors' Instructions}
\titlerunning{Model Checking C Programs with Loops via \textit{k}-Induction and Invariants}

% the name(s) of the author(s) follow(s) next
%
% NB: Chinese authors should write their first names(s) in front of
% their surnames. This ensures that the names appear correctly in
% the running heads and the author index.
%
\author{Herbert Rocha \and Hussama Ismail \and Lucas Cordeiro \and Raimundo Barreto}
%
% \authorrunning{Lecture Notes in Computer Science: Authors' Instructions}
% (feature abused for this document to repeat the title also on left hand pages)

% the affiliations are given next; don't give your e-mail address
% unless you accept that it will be published
\institute{Federal University of Amazonas, Brazil}

%
% NB: a more complex sample for affiliations and the mapping to the
% corresponding authors can be found in the file "llncs.dem"
% (search for the string "\mainmatter" where a contribution starts).
% "llncs.dem" accompanies the document class "llncs.cls".
%

%\toctitle{Lecture Notes in Computer Science}
%\tocauthor{Authors' Instructions}
\maketitle

\begin{abstract}
% \textcolor{red}{
% The first attempts to apply the \textit{k}-induction method to software verification are only recent.~In 
% this paper, we present a novel proof by induction algorithm, which is built
% on the top of a symbolic context-bounded model checker and uses an iterative 
% deepening approach to verify, for each step $k$ up to a given maximum, 
% whether a given safety property $\phi$ holds in the program. 
% The proposed \textit{k}-induction algorithm consists of three different cases called
% \emph{base case}, \emph{forward condition}, and \emph{inductive step}.  
% Intuitively, in the base case, we aim to find a counterexample with up to 
% $k$ loop unwindings; in the forward condition, we check whether loops have 
% been fully unrolled and that $\phi$ holds in all states reachable within $k$ unwindings; 
% and in the inductive step, we check that whenever $\phi$ holds for $k$ unwindings, 
% it also holds after the next unwinding of the system. 
% The algorithm was implemented in two different ways, a sequential and a parallel one, and
% the results were compared. Experimental results show that both forms of the
% algorithm can handle a wide variety of safety properties extracted from standard benchmarks,
% ranging from reachability to time constraints. And by comparison, the parallel algorithm
% solves more verification tasks in less time. This paper marks the first application of the
% \textit{k}-induction algorithm to a broader range of C programs.}
We present a novel proof by induction algorithm, which combines \textit{k}-induction with invariants
to model check C programs with bounded and unbounded loops. 
The \textit{k}-induction algorithm consists of three cases: in the base case, 
we aim to find a counterexample with up to $k$ loop unwindings; in the forward condition, we check whether loops have been 
fully unrolled and that the safety property $P$ holds in all states reachable within $k$ unwindings; 
and in the inductive step, we check that whenever $P$ holds for $k$ unwindings, it also holds after the 
next unwinding of the system. For each step of the \textit{k}-induction 
algorithm, we infer invariants using affine constraints (i.e., polyhedral) to specify pre- and post-conditions.
The algorithm was implemented in two different ways, 
with and without invariants using polyhedral, and the results were compared. Experimental results show that both 
forms can handle a wide variety of safety properties; however, the k-induction 
algorithm adopting polyhedral solves more verification tasks, which demonstrate an improvement of the induction 
algorithm effectiveness.
%\keywords{software engineering, formal methods, bounded model checking, k-induction, invariants.}
\end{abstract}

% % % % % % % % % % % % % % % % % % % % % % % % % % % % % % % % % % % % % % % % % % % % %

% -----------------------------------------------------------------
% => INTRODUCTION -      1 -> [DOING]
% -----------------------------------------------------------------
% 
\input{sections/introduction.tex}

% -----------------------------------------------------------------
% => BACKGROUND        - 2 -> [DOING]
% -----------------------------------------------------------------
% 
\input{sections/background.tex}

% -----------------------------------------------------------------
% EXPERIMENTAL RESULTS - 3 -> [DOING]
% -----------------------------------------------------------------
% 
\input{sections/experimentalresults.tex}
\input{sections/related_work.tex}

% -----------------------------------------------------------------
% CONCLUSIONS AND FUTURE WORKS - 5 -> [DOING]
% -----------------------------------------------------------------
% 
\input{sections/conclusionsandfuturework.tex}

%
%
% BibTeX users please use
% \bibliographystyle{}
% \bibliography{}
%
% Non-BibTeX users please use
%\begin{thebibliography}{}
%
% and use \bibitem to create references.
%
%\bibitem{RefJ}
% Format for Journal Reference
%Author, Journal \textbf{Volume}, (year) page numbers.
% Format for books
%\bibitem{RefB}
%Author, \textit{Book title} (Publisher, place year) page numbers
% etc
%\end{thebibliography}

\end{document}

%% file: sections/introduction.tex
%=-=-=-=-=-=-=-=-=-=-=-=-=-=-=-=-=-=-=-=-=-=-=-=-=-=-=
\section{Introduction}
%=-=-=-=-=-=-=-=-=-=-=-=-=-=-=-=-=-=-=-=-=-=-=-=-=-=-=

% The Bounded Model Checking (BMC) techniques based on Boolean Satisfiability (SAT)~\cite{handbook09}
The Bounded Model Checking (BMC) techniques based on Boolean Satisfiability (SAT)~\cite{handbook09} 
% or Satisfiability Module Theories (SMT)~\cite{BarrettSST09} are successfully applied
or Satisfiability Module Theories (SMT)~\cite{BarrettSST09} are successfully applied
to verify single- and multi-threaded programs and to find 
subtle bugs in real programs~\cite{Clarke04,MerzFS12,Cordeiro12}. 
% subtle bugs in real programs~\cite{Clarke04,MerzFS12,TeseLucas}. 
The idea behind the BMC techniques is to check the negation
of a given property at a given depth, i.e., given a transition system \textit{M},
a property $ \phi $, and a limit of iterations \textit{k}, BMC unfolds the system
\textit{k} times and converts it into a Verification Condition (VC) $ \psi $ such that
$\psi$ is \textit{satisfiable} if and only if $\phi$ has a counterexample of depth
less than or equal to \textit{k}.

Typically, BMC techniques are only able to falsify properties up to 
a given depth \textit{k}; however, they are not able to prove the correctness 
of the system, unless an upper bound of \textit{k} is known, i.e., a bound that 
unfolds all loops and recursive functions to their maximum iteration. 
In particular, BMC techniques limit the size of data structures (e.g., arrays) 
and the number of loop iterations to a given bound $k$. This also 
limits the state space that needs to be explored in software
verification and has allowed BMC tools to find real errors in applications
% \cite{Clarke04,MerzFS12,TeseLucas,Ivancic05}, but at the same time it has also made
\cite{Clarke04,MerzFS12,Cordeiro12,Ivancic05}, but at the same time it has also made 
them susceptible to producing time-out or memory-out for programs that 
contain \emph{unbounded loops} or programs where the number of loop unwindings
cannot be determined statically.  

% Consider for example the simplistic program on the top of
Consider for example the simplistic program on the left of
Fig.~\ref{figure:unwindingprograms} in which the loop in line~2 runs an
unknown number of times, depending on the initial value
non-deterministically assigned to \texttt{x} in line~1. However, the
assertion in line~3 holds independent of \texttt{x}'s initial value.
Unfortunately, BMC tools like CBMC \cite{Clarke04}, LLBMC \cite{MerzFS12}, 
or ESBMC \cite{Cordeiro12} typically fail to verify programs that contain 
such loops. They insert a so-called \emph{unwinding assertion} at the end of the loop,
which consists of the negated loop bound. This enforces BMC tools to choose an 
unwind bound sufficiently large to search deeper in the state space of the program,
but with the drawback of exhausting time and memory resources.
\vspace{-3ex}
\begin{figure}
\centering
\begin{minipage}[t]{0.4\textwidth}
\begin{lstlisting}
unsigned int x=*;
while(x>0) x--;
assert(x==0);
\end{lstlisting}
\end{minipage}
\hspace{1cm}
\begin{minipage}[t]{0.4\textwidth}
\begin{lstlisting}[escapechar=^]
unsigned int x=*;
if (x>0)
  x--; 		^\raisebox{-1pt}[0pt][0pt]{$\Bigg\}\:k\mathrm{~copies}$}^
...     
assert(!(x>0));
assert(x==0);
\end{lstlisting}
\end{minipage}
\caption{Unbounded loop (left) and finite unwinding (right)}
\label{figure:unwindingprograms}
\end{figure}
\vspace{-4ex}

One technique typically used to prove properties, for any given depth, 
is mathematical induction. The algorithm called \textit{k}-induction was successfully 
applied to ensure that (restricted) C programs do not contain data races~\cite{Donaldson10,Kinductor} 
and to respect time constraints specified during the design phase of a system~\cite{EenS03}.
Additionally, the \textit{k}-induction is a well-established technique in hardware verification, 
where it is easier to be applied due to the monolithic transition relation present 
% in hardware designs~\cite{EenS03,GrosseLD09,Sheera00}.~This paper
in hardware designs~\cite{EenS03}.~This paper 
contributes with a new algorithm to prove correctness 
of C programs by mathematical induction in a completely 
automatic way (i.e., the user does not need to provide the loop invariant).

The main idea of the algorithm is to use an iterative deepening approach and check,
for each step \textit{k} up to a maximum value, three different cases called here as base case, forward condition,
and inductive step. Intuitively, in the base case, we intend to find a counterexample of $\phi$ with up to
\textit{k} iterations of the loop. The forward condition checks whether loops have been fully unrolled
and the validity of the property $\phi$ in all states reachable within \textit{k} iterations. 
The inductive step verifies that if $\phi$ is valid for
\textit {k} iterations, then $\phi$ will also be valid for the next unfolding of the system.
For each step of the algorithm, we infer program invariants using affine constraints
to prune the state space exploration and to strength the induction hypothesis.
% in the checking of each case of the algorithm.

% \textcolor{red}{
% The \textit{k}-induction algorithm was implemented in two different ways,
% a sequential and a parallel one, where in the first, the three cases are performed in a sequential manner,
% starting with the base case, followed by the forward condition, and ending with the inductive step.
% In the parallel implementation, three processes are created and monitored by a main process to
% exploit the availablity of multi-core processors in model checking, as done by 
% Holzmann et al.~\cite{HolzmannJG11} and Kahsai et al.~\cite{Kahsai11}.
% Each of the three processes performs a case of the \textit{k}-induction algorithm 
% and communicates the result to the main process.
% }

These algorithms were all implemented in the Efficient SMT-based
Context-Bounded Model Checker tool (known as ESBMC\footnote{Available at http://esbmc.org/}), 
which uses BMC techniques and SMT solvers (e.g.,~\cite{MouraB08,BrummayerB09}) to verify
% embedded systems written in C/C$++$~\cite{Cordeiro12,CordeiroF11}.
embedded systems written in C/C$++$~\cite{Cordeiro12}.
% In Cordeiro et al.~\cite{Cordeiro12,CordeiroFM09,CordeiroFM10} the ESBMC tool is presented, which describes
In Cordeiro et al.~\cite{Cordeiro12} the ESBMC tool is presented, which describes
how the input program is encoded in SMT; 
what the strategies for unrolling loops are; what are the transformations/optimizations that are
important for performance; what are the benefits of using an SMT solver
instead of a SAT solver; and how counterexamples to falsify properties are
reconstructed. Here we extend our previous work and focus our contribution 
on the combination of the \textit{k}-induction algorithm with invariants.  
First, 
we describe the details of an accurate translation that extends ESBMC 
to prove the correctness of a given (safety) property for any depth without manual
annotations of loops invariants.
Second, we adopt program invariants (using polyhedral) in the \textit{k}-induction algorithm, 
to speedup the verification time and to improve the quality 
of the results by solving more verification tasks in less time.
% 
% we use a multi-process implementation
% of the \textit{k}-induction algorithm, similar to Kahsai et al.~\cite{Kahsai11}, 
% to speedup the verification time and to improve the quality 
% of the results by solving more verification tasks in less time. 
%
Third, we show that our present 
implementation is applicable to a broader range of verification tasks, 
where other existing approaches are unable to support~\cite{Donaldson10,Kinductor,GrosseLD09}.

% To validate the implementations of the algorithm, we used the loops \allowbreak benchmarks from the
% International Competition on Software Verification (SV-COMP)~\cite{svcomp2013}.
% 
% The experimental results show that, both the implementations with and without invariants 
% using polyhedral, are able to verify a wide variety of safety properties. 
% The experiments also show that the implementation with invariants presents better results,
% which is able to prove and falsify more properties in the verification tasks, 
% while it requires less verification time.

%% file: sections/background.tex
%=-=-=-=-=-=-=-=-=-=-=-=-=-=-=-=-=-=-=-=-=-=-=-=-=-=-=
\section{Induction-based Verification of C Programs using Invariants}
\label{sec:kinduction}
%=-=-=-=-=-=-=-=-=-=-=-=-=-=-=-=-=-=-=-=-=-=-=-=-=-=-=

The transformations in each step of the \textit{k}-induction algorithm 
take place in the intermediate representation level, after 
converting the C program into a GOTO-program, which simplifies the representation
and handles the unrolling of the loops and the elimination of recursive functions.
% ~For 
% a detailed description of how these simplifications occur for C/C$++$ 
% programs, we refer the reader to Cordeiro et al.~\cite{Cordeiro12} 
% and Ramalho et al.~\cite{ECBS13}.

%=-=-=-=-=-=-=-=-=-=-=-=-=-=-=-=-=-=-=-=-=-=-=-=-=-=-=
\subsection{The Proposed \textit{k}-Induction Algorithm}
\label{sec:k-induction-algorithm}
%=-=-=-=-=-=-=-=-=-=-=-=-=-=-=-=-=-=-=-=-=-=-=-=-=-=-=

Figure~\ref{figure:k-induction-algorithm} shows an overview of the proposed \textit{k}-induction 
algorithm.~We do not add additional details about the transformations on each step of 
the algorithm; we keep it simple and describe the details in the next subsections so that one
can have a big picture of the proposed method. The input of the algorithm is a 
% C/C++ program $P$
C program $P$
together with the safety property $\phi$.~The algorithm returns \textit{true} 
(if there is no path that violates the safety property), \textit{false} (if there exists a path that 
violates the safety property), and \textit{unknown} (if it does not succeed in computing an answer 
\textit{true} or \textit{false}).

In the base case, the algorithm tries to find a counterexample up to a 
maximum number of iterations \textit{k}.~In the forward 
condition, global correctness of the loop w.r.t. the property 
is shown for the case that the loop iterates at most $k$ times;
and in the inductive step, the algorithm checks that, if the property is valid in \textit{k} iterations,
then it must be valid for the next iterations. The algorithm runs up to a maximum number of iterations
and only increases the value of $k$ if it can not falsify the property during the base case.

%=-=-=-=-=-=-=-=-=-=-=-=-=-=-=-=-=-=-=-=-=-=-=-=-=-=-=
\subsubsection{Differences to other \textit{k}-Iduction Algorithms}
%=-=-=-=-=-=-=-=-=-=-=-=-=-=-=-=-=-=-=-=-=-=-=-=-=-=-=

Our \textit{k}-induction algorithm is slightly different than those presented by Gro{\ss}e et al.~\cite{GrosseLD09},
Donaldson et~al.~\cite{Donaldson10}, and Hagen et al.~\cite{Hagen08}.
In Gro{\ss}e et al., the forward condition and the
inductive step are computed differently from our approach (as described in Section~\ref{sec:k-induction-algorithm}) 
and the value of $k$ is increased only at the end of the algorithm; in this particular case, 
computational resources are thus wasted since loops are usually unfolded at least two times.~Donaldson 
et~al.~\cite{Donaldson10} and Hagen et al.~\cite{Hagen08} propose the \textit{k}-induction
with two steps only (i.e., the base case and the inductive step); however, the inductive step
of the approach proposed by Donaldson et~al. requires annotations in the code to introduce loops 
invariants, but our method is completely automatic as in Hagen et al~\cite{Hagen08}.
Additionally, as observed in the experimental evaluation (see Section~\ref{sec:exp}), the use of the forward 
condition, in our proposed method, improves significantly the quality of the results, 
because some programs that are hard to be proved 
% by the inductive step, they are easy to be
by the inductive step can be proved by the forward condition using affine constraints.
\begin{figure}
\centering
\begin{minipage}{0.65\textwidth}
\begin{lstlisting} [escapechar=^]
input: program P and safety property ^$\phi$^
output: true, false, or unknown
k = 1
while k <= max_iterations do
  if base_case(P, ^$\phi$^, k) then
	  show counterexample s[0..k]
    return false
  else
    k=k+1
    if forward_condition(P, ^$\phi$^, k) then
      return true
    else
      if inductive_step(P, ^$\phi$^, k) then
        return true
      end-if
    end-if
  end-if
end-while
return unknown
\end{lstlisting}
\end{minipage}
\caption{An overview of the \textit{k}-induction algorithm.}
\label{figure:k-induction-algorithm}
\end{figure}

%=-=-=-=-=-=-=-=-=-=-=-=-=-=-=-=-=-=-=-=-=-=-=-=-=-=-=
\subsubsection{Loop-free Programs}
\label{sec:multiple-nested-loops}
%=-=-=-=-=-=-=-=-=-=-=-=-=-=-=-=-=-=-=-=-=-=-=-=-=-=-=

In the \textit{k}-induction algorithm, the loop 
unwinding of the program is done incrementally from one 
to \textit{max\_iterations} (cf. Fig.~\ref{figure:k-induction-algorithm}), 
where the number of unwindings is 
% measured by counting the number of \textit{backjumps}~\cite{CFG}.
measured by counting the number of \textit{backjumps}~\cite{CFG}. 
On each step of the \textit{k}-induction algorithm, an instance 
of the program that contains $k$ copies 
of the loop body corresponds to checking a loop-free program, 
which uses only \textit{if}-statements in order to prevent its execution 
in the case that the loop ends before $k$ iterations. 

\begin{mydef}
\label{loop-free-program}
\textbf{(Loop-free Program)}\textit{ A loop-free program is represented by a straight-line program 
(without loops) by providing an $ite\left(\theta, \rho_1, \rho_2\right)$ 
operator, which takes a Boolean formula $\theta$ and, depending on its value, 
selects either the second $\rho_1$ or the third argument $\rho_2$,
where $\rho_1$ represents the loop body and $\rho_2$ represents either another
$ite$ operator, which encodes a \textit{k}-copy of the loop body, or an assertion/assume
statement.}
\end{mydef}

Therefore, each step of our \textit{k}-induction algorithm transforms 
a program with loops into a loop-free program, such that correctness 
of the loop-free program implies correctness of the program with loops.

If the program consists of multiple 
and possibly nested loops, we simply set the number of loop unwindings 
globally, that is, for all loops in the program and apply these aforementioned 
translations recursively. 
% 
% Figure~\ref{figure:iteration-based-unwinding} shows 
% how loop unwindings are applied to a program with nested loops. 
% 
Note, however, that 
each case of the \textit{k}-induction algorithm performs different transformations 
at the end of the loop: either to find bugs (base case) or to prove that enough 
loop unwindings have been done (forward condition).
%
% \begin{figure*}
% \centering
% \includegraphics[scale=0.45]{figuras/iteration-based-unwinding.jpg}
% \caption{(a) A program with nested loops. (b) Iteration-based unwinding of the program in (a).}
% \label{figure:iteration-based-unwinding}
% \end{figure*}

%=-=-=-=-=-=-=-=-=-=-=-=-=-=-=-=-=-=-=-=-=-=-=-=-=-=-=
\subsubsection{Program Translations}
\label{sec:program-translations}
%=-=-=-=-=-=-=-=-=-=-=-=-=-=-=-=-=-=-=-=-=-=-=-=-=-=-=

In terms of program translations, which are all done completely automatic
by our proposed method, the base case simply inserts 
an unwinding assumption, to the respective loop-free program $P'$, consisting 
of the termination condition $\sigma$ after the loop, as follows $I \wedge T \wedge  \sigma \Rightarrow \phi \nonumber$, 
where $I$ is the initial condition, $T$ is the transition
relation of $P'$, and $\phi$ is a safety
property to be checked. 

The forward case inserts an unwinding assertion 
instead of an assumption after the loop, as follows $I \wedge T \Rightarrow \sigma \wedge \phi \nonumber$.
The forward condition, proposed by Gro{\ss}e et al.~\cite{GrosseLD09}, 
introduces a sequence of commands to check whether there is a path between
an initial state and the current state $k$, while in the algorithm
proposed in this paper, an assertion (i.e., the loop invariant) is automatically 
inserted by our algorithm, without the user's intervention, at the end of the loop 
to check whether all states are reached in $k$ steps.
Our base case and forward condition translations can easily be implemented 
on top of plain BMC. 

However, for the inductive step of the algorithm, 
several transformations are carried out. In particular, the loop
$while(c) \left\{ E; \right\}$ is converted into
\begin{equation}
\label{eqwhile-code-transformed}
  \begin{array}{c}
    A; while(c) \left\{ S; E; U; \right\} R;
  \end{array}
\end{equation}
\noindent where $A$ is the code responsible for 
assigning non-deterministic values to all loops variables,
i.e., the state is havocked before the loop, $c$ is the halt condition of the loop \textit{while}, 
$S$ is the code to store the current state of the program variables 
before executing the statements of $E$, $E$ is the actual code inside the 
loop \textit{while}, $U$ is the code to update all program variables
with local values after executing $E$, and $R$ is the code to remove redundant states.
\begin{mydef}
\label{loop-variable}
\textbf{(Loop Variable)}\textit{ A loop variable is a variable $v \subseteq V$, where 
$V = V_{global}\cup V_{local}$ given that $V_{global}$ is the set of global variables 
and $V_{local}$ is the set of local variables that occur in the loop of a program.}
\end{mydef}
\begin{mydef}
\label{havoc-loop-variable}
\textbf{(Havoc Loop Variable)}\textit{ Nondeterministic value is assigned to a 
loop variable $v$ if and only if $v$ is used in the loop termination condition 
$\sigma$, in the loop counter that controls iterations of a loop; 
or repeatedly modified inside the loop body.}
\end{mydef}
The intuitive interpretation of $S$, $U$, and $R$ is that if the current state (after executing $E$)
is different than the previous state (before executing $E$), then new states are produced in the
given loop iteration; otherwise, they are redundant and the code $R$ is then 
responsible for preventing those redundant states to be included into the states vector. 
Note further that the code $A$ assigns non-deterministic values
to all loops variables so that the model checker can explore 
all possible states implicitly. 
Differently, Gro{\ss}e et al.~\cite{GrosseLD09} havoc all program variables, 
which makes it difficult to apply their approach to arbitrary programs
since they do not provide enough information to constrain the havocked 
variables in the program.
Similarly, the loop \textit{for} can easily be converted into the loop \textit{while} as follows:
$for(B; c; D) \left\{ E; \right\}$ is rewritten as
\begin{equation}
\label{eqfor-code-transformedt}
  \begin{array}{c}
    B; \: while(c) \left\{ E; D; \right\}
  \end{array}
\end{equation}

\noindent where $B$ is the initial condition of the loop, $c$ is the halt condition of the loop,
$D$ is the increment of each iteration over $B$, and $E$ is the actual code inside the loop \textit{for}. 
No further transformations are applied to the loop \textit{for} during the inductive step.
Additionally, the loop \textit {do while} can trivially be 
converted into the loop \textit{while} with one difference, 
the code inside the loop must execute
at least once before the halt condition is checked.

The inductive step is thus represented by
$\gamma \wedge \sigma \Rightarrow \phi$,
where $\gamma$ is the transition relation of $\hat{P}$, which represents 
a loop-free program (cf. Definition~\ref{loop-free-program}) after applying 
transformations (\ref{eqwhile-code-transformed}) and (\ref{eqfor-code-transformedt}).
The intuitive interpretation of the inductive step is to prove that,
for any unfolding of the program, there is no assignment of particular 
values to the program variables that violates the safety property being checked.
Finally, the induction hypothesis of the inductive step consists of the 
conjunction between the postconditions ($Post$) and the 
termination condition ($\sigma$) of the loop.

%=-=-=-=-=-=-=-=-=-=-=-=-=-=-=-=-=-=-=-=-=-=-=-=-=-=-=
\subsubsection{Invariant Generation}
\label{sec:invariant-gen}
%=-=-=-=-=-=-=-=-=-=-=-=-=-=-=-=-=-=-=-=-=-=-=-=-=-=-=

To infer program invariants, we adopted the PIPS~\cite{Maisonneuve:2014} tool, which 
is an interprocedural source-to-source compiler framework for C and Fortran programs and 
relies on a polyhedral abstraction of program behavior. 
PIPS development has been driven for almost twenty years to automatic analysis of large size 
programs~\cite{PIPS:2015}. PIPS performs a two-step analysis: (1) each program instruction is associated to an affine transformer, 
representing its underlying transfer function. This is a bottom-up procedure, starting from elementary instructions, 
then working on compound statements and up to function definitions; (2) polyhedral invariants are propagated along 
with instructions, using previously computed transformers.

% In our proposed method, PIPS receives as input the analyzed program and then it generates
In our proposed method, PIPS receives the analyzed program as input and then it generates 
invariants that are given as comments surrounding instructions in the output C code. 
These invariants are translated and instrumented into the program as assume statements. 
In particular, we adopt the function \texttt{assume($expr$)} 
to limit possible values of the variables that are related to the invariants.
This step is needed since PIPS generates invariants that are presented as mathematical expressions 
(e.g., $2j < 5t$), which are not accepted by C programs syntax and invariants with $\#init$ 
suffix that is used to distinguish the old value from the new value. 

%=-=-=-=-=-=-=-=-=-=-=-=-=-=-=-=-=-=-=-=-=-=-=-=-=-=-=
\subsection{Running Example}
\label{sec:running-example}
%=-=-=-=-=-=-=-=-=-=-=-=-=-=-=-=-=-=-=-=-=-=-=-=-=-=-=

% As a running example, a program extracted from the benchmarks 
% of the SV-COMP~\cite{svcomp2013} is used as shown in Figure~\ref{figure:codigo-main},
% 
A program extracted from the benchmarks of the SV-COMP~\cite{svcomp2013} is used as a 
running example as shown in Figure~\ref{figure:codigo-main}, 
which already includes invariants using polyhedral. 
 \begin{figure}
 \centering
 {\tiny
 \begin{minipage}[t]{0.8\textwidth}
 \begin{lstlisting}[escapechar=^, escapeinside={@}{@}]
 int main(int argc, char **argv)
{ 
 long long int i = 1, sn = 0; 
 assume( i==1 && sn==0 ); // Invariant
 unsigned int n; 
 assume(n>=1);
 while (i<=n) {  
   assume( 1<=i && i<=n ); // Invariant
   sn = sn+a;  
   i++;
 }
 assume( 1<=i && n+1<=i ); // Invariant
 assert(sn==n*a); @\label{codeex:assert}@
}
 \end{lstlisting}
 \end{minipage}
 }
\caption{Running example for the \textit{k}-induction algorithm.}
\label{figure:codigo-main}
\end{figure}
%
% \begin{figure}
% \centering
% \begin{minipage}{0.6\textwidth}
% \begin{lstlisting}
% int main(int argc, char **argv) {
%   long long int i=1, sn=0;
%   unsigned int n;
%   assume (n>=1);
%   while (i<=n) {
%     sn = sn + a;
%     i++;
%   }
%   assert(sn==n*a);
% }
% \end{lstlisting}
% \end{minipage}
% \caption{Running example for the \textit{k}-induction algorithm.}
% \label{figure:codigo-main}
% \end{figure}
%
In Figure~\ref{figure:codigo-main}, $a$ is an integer constant and
note that variables $i$ and $ sn$ are declared with a type larger than 
the type of the variable $n$ to avoid arithmetic overflow. 
Mathematically, the code above represents the implementation
of the sum given by the following equation:
\begin{equation}
\label{equ:soma-induction}
 S_n = \sum_{i=1}^n a = na, n \geq 1
\end{equation}
In the code of Figure~\ref{figure:codigo-main}, the invariants produced 
by PIPS are included as assume statements; the property
(represented by the assertion in line $\ref{codeex:assert}$) must be \textit{true}
for any value of \textit{n} (i.e., for any unfolding of 
the program). 
\comment{
This code can be modeled by a transition system. 
% , as shown in Figure~\ref{transition-code-main}.
The initial state represents the code before the loop, which contains
the initialization of variables and assumptions about them.
The following two states represent the code that executes in the loop,
i.e., the computation of the variable $sn$ and the increment of the
variable $i$. A transition $ \tau $ represents the immediate change of state,
in this case, after calculating $sn$. The final state
is the code after the loop, which contains an assertion about the value
of the variable $sn$.}
Differently from our \textit{k}-induction algorithm,
BMC techniques have difficulties in proving the correctness
of this (simple) program since the upper limit value of the loop, represented by $n$, 
is non-deterministically chosen (i.e., the variable $n$ can assume 
any value from one to the size of the \textit{unsigned int} type, 
which varies between different types of computers).
Due to this condition, the loop will be unfolded $2^n-1$ times
(in the worst case, $2^{32} - 1$ times on $32$ bits integer), which is thus impractical. 
Basically, the bounded model checker would symbolically execute several times
the increment of the variable $i$ and the computation of the variable $sn$
by $4,294,967,295$ times. To solve the problem of unfolding the loop $2^n-1$ times,
the translations previously described are performed.
%
% \begin{figure}
%  \centering
%  \input{figuras/sum01.tex}
%  \caption{The state transition system of the code shown in Figure~\ref{figure:codigo-main}.}
%  \label{transition-code-main}
% \end{figure}
%
\comment{
In the next subsections, we explain how the \textit{k}-induction algorithm 
(see Figure~\ref{figure:k-induction-algorithm}) can prove correctness of the C program
shown in Figure~\ref{figure:codigo-main}.}

%=-=-=-=-=-=-=-=-=-=-=-=-=-=-=-=-=-=-=-=-=-=-=-=-=-=-=
\subsubsection{The Base Case}
\label{sec:base-case}
%=-=-=-=-=-=-=-=-=-=-=-=-=-=-=-=-=-=-=-=-=-=-=-=-=-=-=

The base case initializes the limits of the loop's termination 
condition with non-deterministic values so that the 
model checker can explore all possible states implicitly. 
The pre- and postconditions of the loop
shown in Figure~\ref{figure:codigo-main}, in static single assignment (SSA) form~\cite{CFG}, are as follows:
\begin{equation}
\label{base-case-p}
Pre := \left [ \begin{array}{ll} 
                n_1 = nondet\_uint \wedge n_1 \geq 1   \\
                \wedge \, \, sn_1 = 0 \wedge i_1 = 1 \\ 
              \end{array} \right ]  \nonumber \\ 
\end{equation}
\begin{equation}
\label{base-case-r}
Post := \left [ \begin{array}{ll} 
                i_k \geq 1 \wedge \, \, i_k > n_1 \Rightarrow sn_{k} = n_1 \times a  \\
              \end{array} \right ]  \nonumber \\ 							
\end{equation}

\noindent where $Pre$ and $Post$ are the pre- and postconditions to compute the sum
given by Equation (\ref{equ:soma-induction}), respectively,
and $nondet\_uint$ is a non-deterministic function, which can return any value
of type \textit{unsigned int}.~In the preconditions, $n_1$ represents the first
assignment to the variable $n$, which is a non-deterministic value greater than or equal to one.
This ensures that the model checker explores all possible unwindings of the program.~Additionally, 
$sn_1$ represents the first assignment to the variable $sn$ 
and $i_1$ is the initial condition of the loop.
In the postconditions,  $sn_{k}$ represents the assignment $n+1$ for
the variable $sn$ in Figure~\ref{figure:codigo-main},
which must be \textit{true} if $i_k> n_1$. The code that is not pre- or postcondition
is represented by the variable $Q$ (i.e., the sequence of commands inside the loop \textit{for}) 
and it does not undergo any transformation during the base case.
The resulting code of the base case transformations
can be seen in Figure~\ref{figure:codigo-main-caso-base} (cf. Definition~\ref{loop-free-program}). Note that the \textit{assume}
(in line $11$), which consists of the termination condition, eliminates all
execution paths that do not satisfy the constraint $i > n$. 
This ensures that the base case finds a counterexample of depth $k$
without reporting any false negative result.
Note further that other assume statements, shown in Figure~\ref{figure:codigo-main},
are simply eliminated during the symbolic execution by propagating constants and 
checking that the resulting expression evaluates to \textit{true}~\cite{Cordeiro12}.

\begin{figure}
\centering
\begin{minipage}{0.8\textwidth}
\begin{lstlisting} [escapechar=^]
int main(int argc, char **argv) {
  long long int i, sn=0;
  unsigned int n=nondet_uint();
	assume (n>=1);
  i=1;
  if (i<=n) {
    sn = sn + a; ^\raisebox{-1pt}[0pt][0pt]{$\Bigg\}\:k\mathrm{~copies}$}^
    i++;
  }
	...
  assume(i>n); // unwinding assumption
  assert(sn==n*a);
}
\end{lstlisting}
\end{minipage}
\caption{Example code for the proof by mathematical induction, during base case.}
\label{figure:codigo-main-caso-base}
\end{figure}

\vspace{-5ex}
%=-=-=-=-=-=-=-=-=-=-=-=-=-=-=-=-=-=-=-=-=-=-=-=-=-=-=
\subsubsection{The Forward Condition}
\label{sec:forward-condition}
%=-=-=-=-=-=-=-=-=-=-=-=-=-=-=-=-=-=-=-=-=-=-=-=-=-=-=

In the forward condition, the \textit{k}-induction algorithm attempts to prove that the
loop is sufficiently unfolded and whether the property is valid in all states reachable
within \textit{k} steps. The postconditions of the loop shown in
Figure~\ref{figure:codigo-main}, in SSA form, can then be defined as follows:
\comment{
\begin{equation}
\label{forward-case-p}
Pre := \left [ \begin{array}{ll} 
                n_1 = nondet\_uint \wedge n_1 \geq 1   \\
                \wedge \, \, sn_1 = 0 \wedge i_1 = 1  \\ 
              \end{array} \right ]  \nonumber \\ 
\end{equation}}
\begin{equation}
\label{forward-case-r}
Post := \left [ \begin{array}{ll} 
                i_{k} > n_1 \wedge sn_{k} = n_1 \times a  \\
              \end{array} \right ]  \nonumber \\ 							
\end{equation}

The preconditions of the forward condition are identical to the base case.
In the postconditions $Post$, there is an assertion to check whether
the loop is sufficiently expanded, represented by the constraint $i_{k} > n_1$,
where $ i_{k} $ represents the value of the variable $i$ at iteration $n + 1$. 
The resulting code of the forward condition transformations 
can be seen in Figure~\ref{figure:codigo-main-forward-condition} (cf. Definition~\ref{loop-free-program}).
The forward condition attempts to prove that the loop is unfolded deep enough
(by checking the loop invariant in line $11$) and if the property is valid in all states reachable
within \textit {k} iterations (by checking the assertion in line $12$).
As in the base case, we also eliminate assume expressions by checking whether they evaluate to \textit{true}
by propagating constants during symbolic execution.

\begin{figure}
\centering
\begin{minipage}{0.8\textwidth}
\begin{lstlisting} [escapechar=^]
int main(int argc, char **argv) {
  long long int i, sn=0;
  unsigned int n=nondet_uint();
	assume (n>=1);
  i=1;
  if (i<=n) {
    sn = sn + a; ^\raisebox{-1pt}[0pt][0pt]{$\Bigg\}\:k\mathrm{~copies}$}^
    i++;
  }
	...
  assert(i>n); // check loop invariant
  assert(sn==n*a);
}
\end{lstlisting}
\end{minipage}
\caption{Example code for the proof by mathematical induction, during forward condition.}
\label{figure:codigo-main-forward-condition}
\end{figure}

%=-=-=-=-=-=-=-=-=-=-=-=-=-=-=-=-=-=-=-=-=-=-=-=-=-=-=
\subsubsection{The Inductive Step}
\label{sec:inductive-step}
%=-=-=-=-=-=-=-=-=-=-=-=-=-=-=-=-=-=-=-=-=-=-=-=-=-=-=

In the inductive step, the \textit{k}-induction algorithm attempts to prove that,
if the property is valid up to depth $k$, the same must be valid for the next value of $k$.~Several 
changes are performed in the original code during this step. First,
a structure called \textit{statet} is defined, containing all variables within the loop and
the halt condition of that loop. Then, a variable of type \textit{statet} called $cs$ (current state)
is declared, which is responsible for storing the values of a given variable in a given iteration;
in the current implementation, the $cs$ data structure does not handle heap-allocated objects.
A state vector of size equal to the number of iterations of the loop is also declared, 
called $sv$ (state vector) that will store the values of all variables on each iteration of the loop.

Before starting the loop, all loops variables (cf. Definitions~\ref{loop-variable} and~\ref{havoc-loop-variable}) 
are initialized to non-deterministic values
and stored in the state vector on the first iteration of the loop so that the 
model checker can explore all possible states implicitly.
Within the loop, after storing the current state and executing the code inside the loop,
all state variables are updated with the current values of the current iteration.
An \textit{assume} instruction is inserted with the condition that the current state is
different from the previous one, to prevent redundant states to be inserted into the state vector;
in this case, we compare $sv_{j}\left[i\right]$ to $cs_{j}$ for $0 < j \leq k$ and $0 \leq i < k$.
In the example we add constraints as follows:
\begin{equation} 
\begin{split}
&sv_{1}\left[0\right] \neq cs_{1} \\ 
&sv_{1}\left[0\right] \neq cs_{1} \wedge sv_{2}\left[1\right] \neq cs_{2} \\ 
&\cdots \\ 
&sv_{1}\left[0\right] \neq cs_{1} \wedge sv_{2}\left[1\right] \neq cs_{2} \wedge \ldots sv_{k}\left[i\right] \neq cs_{k} \\
\end{split}
\label{eq:inductive-step-constraints}
\end{equation} 
Although, we can compare $sv_{k}\left[i\right]$ to all $cs_{k}$ for $i<k$ (since 
inequalities are not transitive), the number of constraints 
can still grow very large quickly, and easily ``blow-up'' the SMT solver.
In the SV-COMP benchmarks, we observed a substantial improvement in performance 
if we generate and check constraints as described in Equation (\ref{eq:inductive-step-constraints}).

% Finally, after the loop an \textit{assume} instruction is inserted, which is similar to that inserted
Finally, an \textit{assume} instruction is inserted after the loop, which is similar to that inserted
in the base case. The pre- and postconditions of the loop shown in Figure~\ref{figure:codigo-main},
in SSA form, are defined as follows:
\begin{equation}
\label{inductive-step-p}
Pre := \left [ \begin{array}{ll} 
                n_1 = nondet\_uint \wedge n_1 \geq 1   \\
								\wedge \, \, sn_1 = 0 \wedge i_1 = 1 \\
                \wedge \, \, cs_1.v_0=nondet\_uint  \\
								\wedge \, \, \ldots \\
								\wedge \, \, cs_1.v_m=nondet\_uint \\
              \end{array} \right ]  \nonumber \\ 
\end{equation}
\begin{equation}
\label{inductive-step-r}
Post := \left [ \begin{array}{ll} 
                i_k > n_1 \Rightarrow sn_{k} = n_ \times a \\
              \end{array} \right ]  \nonumber \\ 							
\end{equation}

In the preconditions $Pre$, in addition to the initialization of the variables,
the value of all variables contained in the current state $cs$ must be assigned
with non-deterministic values, where $m$ is the number of (automatic and static) variables
that are used in the program. The postconditions do not change, as in the base case;
they only contain the property that the algorithm is trying to prove.
In the instruction set $Q$, changes are made in the code to save the value of
the variables before and after the current iteration $i$, as follows:
\begin{equation}
\label{inductive-step-q}
Q := \left [ \begin{array}{ll} 
                sv[i-1] = cs_i \wedge S  \\
                \wedge \, \,  cs_i.v_0=v_{0i}  \\
								\wedge \, \, \ldots \\
								\wedge \, \, cs_i.v_m=v_{mi} \\
              \end{array} \right ]  \nonumber \\ 
\end{equation}

In the instruction set $Q$, $sv [i-1]$ is the vector position to save the current state
$cs_i$, $S$ is the actual code inside the loop, and the assignments 
$cs_i.v_0=v_{0i} \: \wedge \: \ldots \: \wedge \: cs_i.v_m=v_{mi}$
represent the value of the variables in iteration $i$ being saved in the current state $cs_i$.~The 
modified code for the inductive step, using the notation defined in Section~\ref{sec:k-induction-algorithm}, 
can be seen in Figure~\ref{figure:codigo-main-passo-indutivo}. 
Note that the \textit{if}-statement (lines~18-26)
in Figure~\ref{figure:codigo-main-passo-indutivo} is copied \textit{k}-times according to Definition~\ref{loop-free-program}.
As in the base case, the inductive step also inserts an \textit{assume} 
instruction, which contains the termination condition.~Differently from base case, 
the inductive step proves that the property, specified 
by the assertion, is valid for any value of $n$.
\begin{mylem}
\label{lemma:forward-condition}
\textit{If the induction hypothesis $\{Post \: \: \wedge \: \: \neg \left(i \leq n\right)\}$ 
holds for $k+1$ consecutive iterations, then it also holds for $k$ preceding iterations.}
\end{mylem}
\noindent After the loop \textit{while} is finished, the induction hypothesis 
$\{Post \: \: \wedge \: \: \neg \left(i \leq n\right)\}$ 
is satisfied on any number of iterations; in particular, the SMT solver 
can easily verify Lemma~\ref{lemma:forward-condition} 
and conclude that $sn==n*a$ is inductive relative to $n$.
As in previous cases, we also eliminate assume expressions by checking whether they evaluate to \textit{true}
by propagating constants during symbolic execution.
\begin{figure}[t]
\centering
\begin{minipage}{0.7\textwidth}
\begin{lstlisting}
//variables inside the loop
typedef struct state {
 long long int i, sn;
 unsigned int n;
} statet;
int main(int argc, char **argv) {
 long long int i, sn=0;
 unsigned int n=nondet_uint();
 assume (n>=1);
 i=1;
 //declaration of current state
 //and state vector
 statet cs, sv[n];
 //A: assign non-deterministc values
 cs.i=nondet_uint(); 
 cs.sn=nondet_uint();
 cs.n=n;
 if (i<=n) { //c: halt condition
  sv[i-1]=cs;  //S: store current state
  sn = sn + a; //E: code inside the loop
  //U: update variables with local values
  cs.i=i; cs.sn=sn; cs.n=n;
  //R: remove redundant states
  assume(sv[i-1]!=cs);
  i++;
 }
 ...
 assume(i>n); //unwinding assumption
 assert(sn==n*a);
}
\end{lstlisting}
\end{minipage}
\caption{Example code for the proof by mathematical induction, during inductive step.}
\label{figure:codigo-main-passo-indutivo}
\end{figure}

%% file: sections/experimentalresults.tex
%=-=-=-=-=-=-=-=-=-=-=-=-=-=-=-=-=-=-=-=-=-=-=-=-=-=-=
\section{Experimental Evaluation}
\label{sec:exp}
%=-=-=-=-=-=-=-=-=-=-=-=-=-=-=-=-=-=-=-=-=-=-=-=-=-=-=

To evaluate the proposed method, we initially adopted the benchmarks of the 
SV-COMP $2015$\footnote{http://sv-comp.sosy-lab.org/2015/}, in particular the \textit{Loops} subcategory. 
The \textit{k}-induction algorithm was implemented in two different ways, 
with and without invariants using polyhedral. 
% , and the results were compared. 
The implementation of the algorithm with invariants is called DepthK~\footnote{https://github.com/hbgit/depthk}. 
% The ESBMC v$1.24.1$ was adopted in the both implementation.
The ESBMC v$1.24.1$ was adopted in both implementation. 
This way, we performed a comparison between 
DepthK (i.e., \textit{k}-induction and invariants), 
ESBMC with \textit{k}-induction, 
and ESBMC with plain BMC.

\subsection{Experimental Setup}

The experiments were conducted on a computer with Intel Core i7-$2600$, $3.40$GHz with 
$24$GB of RAM with Ubuntu $14.04.1$ LTS $64$-bit. 
Each verification task is limited to a CPU run time of $15$ min and a 
memory consumption of $15$ GB. Additionally, we defined the \textit{max\_iterations} 
to $100$ (cf. Fig.~\ref{figure:k-induction-algorithm}). 
% 
% The tool used,
% with the implementation of the \textit{k}-induction algorithm, is ESBMC v1.24.1
% ~\cite{MorseCNF13}.
% Although ESBMC supports incremental SMT solving, it did not work well in practice for the
% \textit{k}-induction algorithm since the overhead of repeatedly checking the formulae 
% satisfiability outweighs the amount of state space that is pruned by 
% reusing previously learned clauses.

\textit{Loops} subcategory consists of $142$ verification tasks, which are organized as follows: 
$49$ benchmarks contain valid properties (i.e., the verification tool must be able to prove correctness) and 
$93$ benchmarks contain invalid properties (i.e., the verification tool must be able to falsify the property).

%=-=-=-=-=-=-=-=-=-=-=-=-=-=-=-=-=-=-=-=-=-=-=-=-=-=-=
\subsection{Experimental Results}
\label{results-kinduction}
%=-=-=-=-=-=-=-=-=-=-=-=-=-=-=-=-=-=-=-=-=-=-=-=-=-=-=

\begin{figure}
\centering
\begin{minipage}{.5\textwidth}
  \centering
  \includegraphics[width=1\linewidth]{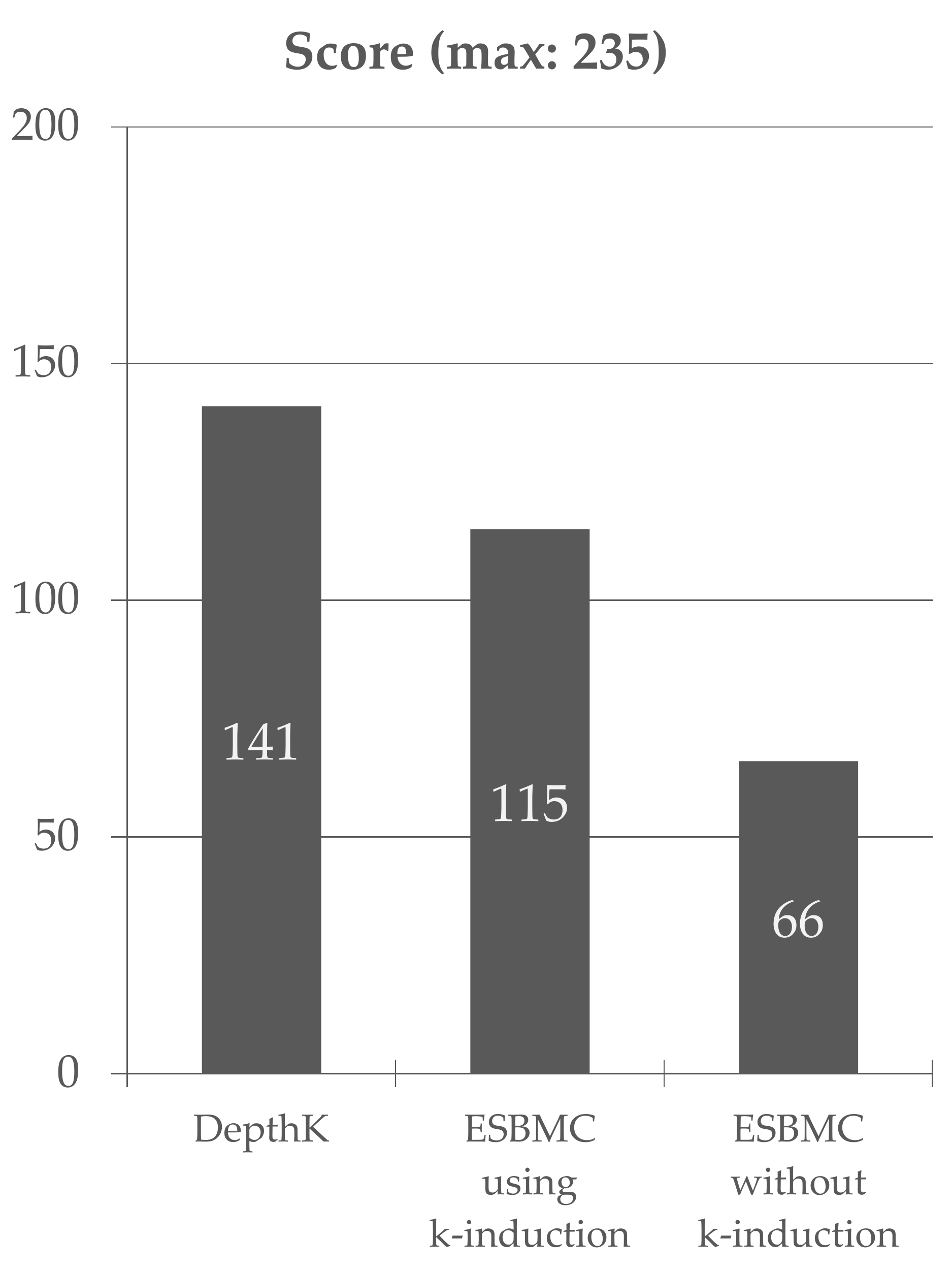}
  \captionof{figure}{Verification Scores}
  \label{figure:k-score}
\end{minipage}%
\begin{minipage}{.5\textwidth}
  \centering
  \includegraphics[width=1\linewidth]{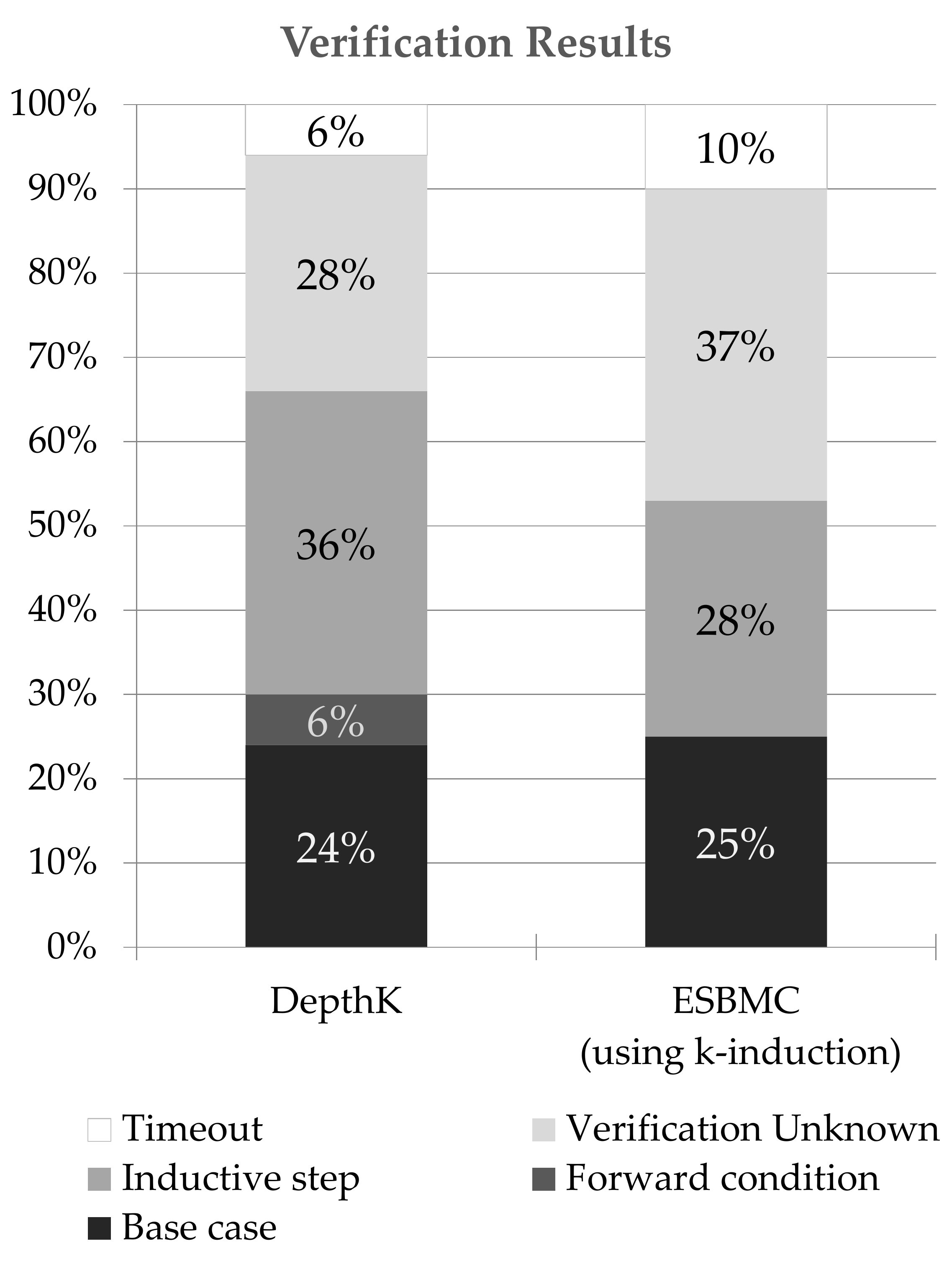}
  \captionof{figure}{Verification results}
  \label{figure:k-results}
\end{minipage}
\end{figure}

We evaluate the experimental results as follows: 
we adopt the same score scheme that is used by the SVCOMP rules\footnote{http://sv-comp.sosy-lab.org/2015/rules.php};  
in particular, we check the verification result and time presented by each implementation. 
Figure~\ref{figure:k-score} shows the comparative results between the scores generated by 
DepthK with \textit{k}-induction and invariants using polyhedral, ESBMC using k-induction only, 
and ESBMC using plain BMC.
The total scores in the \textit{Loops} subcategory for
ESBMC with plain BMC is $66$; 
ESBMC using k-induction only is $115$; and  
DepthK combining \textit{k}-induction and invariants is $141$.

% \begin{figure*}
%   \centering
%   \includegraphics[scale=0.44]{graphs/results}
%     \caption{\textcolor{blue}{Verification results for \textit{Loops} Category of SV-COMP15 detailed by steps for DepthK, ESBMC using \textit{k}-induction and ESBMC without  \textit{k}-induction approach.}}
%   \label{figure:k-results}
% \end{figure*}

Figure~\ref{figure:k-results} shows the distribution of the result by each step of the \textit{k}-induction algorithm 
(i.e., base case, forward condition, and inductive step), 
including verifications that result in unknown and timeout. 
If we analyze the distribution of the results,
% 
% We identify that 
% DepthK has been find violations in $24$\% of the benchmarks in the base case, it ensures 
% valid properties for $6$\% in forward condition, and $36$\% in inductive step. 
% % 
% ESBMC adopting only \textit{k}-induction has been find  
% $25$\% of violations in the base case, while proving correctness was done only by inductive step in $28$\% of the 
% benchmarks. 
% 
% TODO: Talk how the forward condition is important
we identified that DepthK was able to prove properties during the forward condition in $6$\% of the verification tasks, and 
ESBMC with \textit{k}-induction proves properties only during the inductive step. As a result, we observe that invariants 
help prove that the loop is sufficiently unfolded and whether the property is valid, taking into account that this is performed 
in the forward condition step.
DepthK has not found a solution in $28$\% of the verification tasks; this is explained by the invariants generated from PIPS, which  
could not generate strong invariants to be k-inductive either due to the transformers or due to the invariants are not convex.
ESBMC with \textit{k}-induction does not find a solution in $37$\% of the verification tasks, therefore providing 
evidences that invariants can improve the verification results. 
DepthK and ESBMC with \textit{k}-induction found a solution in $66$\% and $53$\% of the verification tasks, respectively. However,
they reported $6$\%  and $10$\% of timeouts, respectively.

% The \textit{k}-induction verification is split into three steps (i.e., base case, forward condition, and inductive step). 

% \begin{figure*}
%   \centering
%   \includegraphics[scale=0.44]{graphs/time}
%     \caption{\textcolor{blue}{Verification time in seconds for \textit{Loops} Category of SV-COMP15 by DepthK, ESBMC using \textit{k}-induction and ESBMC without  \textit{k}-induction approach.}}
%   \label{figure:k-time}
% \end{figure*}

% As shown in Figure~\ref{figure:k-time}, 
Analyzing the verification time, we identified that 
ESBMC with \textit{k}-induction verified all benchmarks in $15320$ seconds;  
DepthK took approximately $11518$ seconds, i.e., a reduction of $25$\% of the verification time;  
and ESBMC using plain BMC took about $3100$ seconds (but solves less verification tasks).

%% file: sections/related_work.tex
\section{Related Work}
\label{sec:related}
%=-=-=-=-=-=-=-=-=-=-=-=-=-=-=-=-=-=-=-=-=-=-=-=-=-=-=

The application of the \textit{k}-induction method is gaining popularity 
in the software verification community. 
Recently, Bradley et al. introduce
 ``property-based reachability'' (or IC$3$) procedure for the safety 
verification of systems~\cite{Bradley12,Bradley13}. The authors have 
shown that IC$3$ can scale on certain benchmarks where \textit{k}-induction 
fails to succeed. However, we do not compare \textit{k}-induction against
IC$3$ since it is already done by Bradley~\cite{Bradley12}; we focus
our comparison on related \textit{k}-induction procedures.

Previous work on the one hand 
have explored proofs by mathematical induction of hardware and software 
systems with some limitations, e.g., requiring changes in the code 
to introduce loop invariants~\cite{Donaldson10,Kinductor,GrosseLD09}.~This 
complicates the automation of the verification process,
unless other methods are used in combination to automatically compute the loop 
invariant~\cite{SharmaDDA11,AncourtCI10}. 
Similar to the approach proposed by Hagen and Tinelli~\cite{Hagen08}, 
our method is completely automatic and does not require the user to provide 
loops invariants as the final assertions after each loop.
On the other hand, state-of-the-art BMC tools have been widely used, 
but as bug-finding tools since they typically analyze 
bounded program runs~\cite{Clarke04,MerzFS12}; completeness can only be ensured 
if the BMC tools know an upper bound on the depth of the state space, 
which is not generally the case.~This paper closes this gap, 
providing clear evidence that the \textit{k}-induction algorithm can be applied to a broader
range of C programs without manual intervention.

Gro{\ss}e et al. describe a method to prove properties of TLM designs 
(Transaction Level Modeling) in SystemC~\cite{GrosseLD09}. 
The approach consists of converting a SystemC program into a C program, 
and then it performs the proof of the properties by mathematical induction 
using the CBMC tool~\cite{Clarke04}. 
% 
% The approach consists of three steps; starting with the transformation 
% of a SystemC program into a C program, followed by the generation 
% and addition of logics to monitor TLM properties (using assertions 
% and finite state machines), and the verification of the C program 
% using the CBMC tool~\cite{Clarke04}.
% ~In CBMC, the proof of the properties 
% is done by mathematical induction, using the \textit{k}-induction algorithm
% implemented by the authors. 
% A producer-consumer program is tested, 
% however, with various restrictions. The tool is not able to finalize 
% the verification in the given time with the increase of the number 
% of producers and consumers; additionally, the technique can only prove correctness 
% for a limited number of iterations. 
% The \textit{k}-induction algorithm, implemented 
% by the authors, is similar to the one described in this paper.  
% which uses three steps: base case, forward condition, and inductive step. 
%
The difference to the one described in this paper lies on the transformations carried out in the forward condition. 
% In the \textit{k}-induction algorithm, proposed by Gro{\ss}e et al., during the forward condition,
During the forward condition, 
transformations similar to those inserted during the inductive step 
in our approach, are introduced in the code to check whether 
there is a path between an initial state and the current state $k$; 
while the algorithm proposed in this paper, an assertion is inserted 
at the end of the loop to verify that all states are reached in $k$ steps. 

Donaldson et~al.\ describe a verification tool called Scratch~\cite{Donaldson10} 
to detect data races during Direct Memory Access (DMA) in the CELL BE processor 
from IBM~\cite{Donaldson10}. The approach used to verify C programs is the 
\textit{k}-induction technique. 
% The tool inserts assertions into the program to model 
% the behavior of the memory control-flow, and it attempts to prove the correctness 
% of the program, using the \textit{k}-induction algorithm. 
The approach was implemented in the Scratch tool that uses two steps, the base case and the inductive step.
The tool is able to prove the absence of data races, but it is 
restricted to verify that specific class of problems for a particular type of hardware. 
% Differently from the algorithm proposed in this paper, Donaldson et al. proposed the \textit{k}-induction 
% with two steps, the base case and the inductive step. 
The steps of the algorithm are similar to the one proposed in this paper, but it requires annotations in the code
to introduce loops invariants. 
Kahsai et al. describe PKIND, a parallel version of the tool KIND, 
used to verify invariant properties of programs written in Lustre~\cite{Kahsai11}. 
% PKIND makes use of a multi-process approach, 
% , similar to ESBMC, with the difference that 
% the communication between process is message-based for PKIND (using MPI API~\cite{MPI}), 
% the communication between process is message-based for PKIND.
% and pipe-based for ESBMC. 
In order to verify a Lustre program, PKIND starts three processes, 
one for base case, one for inductive step, and one for invariant generation, 
note that unlike ESBMC, the k-induction algorithm used by PKIND does not have a forward 
condition step. The base case starts the verification with $k=0$, and increments its value 
until it finds a counterexample or it receives a message from the inductive step process that 
a solution was found. 
Similarly, the inductive step increases the value 
of $k$ until it receives a message from the base case process or a solution is found. 
The invariant generation process generates a set of candidates invariants from predefined 
templates and constantly feeds the inductive step process, as done recently by Beyer et al.~\cite{Beyer15}	
(we do not compare to Beyer et al. since their technical report appeared only
after we submitted our CAV abstract and thus there was no time to further evaluate their work).
% It is important to note 
% that when the inductive step finds a solution for an arbitrary value of $k$, it sends a message 
% to the base case process that checks if the properties holds up to that value, before showing 
% the result.

%% file: sections/conclusionsandfuturework.tex
%=-=-=-=-=-=-=-=-=-=-=-=-=-=-=-=-=-=-=-=-=-=-=-=-=-=-=
\section{Conclusions}
\label{sec:conc}
%=-=-=-=-=-=-=-=-=-=-=-=-=-=-=-=-=-=-=-=-=-=-=-=-=-=-=

The main contributions of this work are the design, implementation, and evaluation of the 
\textit{k}-induction algorithm adopting invariants using polyhedral in a verification tool, 
as well as, the use of the technique for the automated verification of reachability properties 
in heteregenous programs. 
To the best of our knowledge, this paper marks the first application of the \textit{k}-induction algorithm 
to a broader range of C programs with loops.
To validate the \textit{k}-induction algorithm, experiments were performed 
involving $142$ benchmarks of the SV-COMP 2015 \textit{loops} subcategory. 
The experimental results also show that the \textit{k}-induction algorithm without invariants was able 
to verify $52\%$ of the benchmarks in $15320$ seconds, and \textit{k}-induction algorithm with invariants 
using polyhedral was able to verify $66\%$ of the benchmarks in $11518$ seconds, which gives a speedup of 
roughly $25\%$ faster than the \textit{k}-induction algorithm without the invariants version.
Given a fixed timeout, this speedup can also 
improve the quality of the results (around $13\%$), because more programs 
can be verified if their verification would otherwise be interrupted by the time limit. 
% In addition, the algorithm was able to prove and falsify safety properties and temporal constraints 
% in a bicycle computer. 
In addition, both forms were able to prove or falsify a wide variety of safety properties; 
however, the \textit{k}-induction algorithm adopting polyhedral solves more 
verification tasks, which demonstrate an improvement of the induction algorithm effectiveness.

% In the two proposed approaches, the parallel implementation of the \textit{k}-induction 
% produced better results, with the largest number of benchmarks being successfully verified in less time. 

% For future work, we intend to investigate whether redundant computations (or constraints) 
% in the \textit{k}-induction algorithm can be avoided, possibly by using the results 
% of already completed steps.
% using the Green solver interface~\cite{Visser12}.